\documentclass[journal]{IEEEtran}
\usepackage{algpseudocode}
\usepackage[T1]{fontenc}
\usepackage{amsmath}
\usepackage{amsthm,amssymb,amsmath,bm}
\usepackage{subfigure}
\usepackage{amsfonts}
\usepackage{epsfig}
\usepackage{amssymb}
\usepackage{amsmath}
\usepackage[table,xcdraw]{xcolor}
\usepackage{cite}
\usepackage[utf8x]{inputenc}
\usepackage{color,soul}
\usepackage{subfigure}
\usepackage{multirow}
\usepackage{rotating}
\usepackage{graphicx}
\usepackage{tabularx}
\usepackage{array}
\usepackage{color,soul}
\usepackage{bm}
\usepackage{graphicx,dblfloatfix}
\usepackage{blindtext}

\usepackage{subfigure}
\usepackage{moreverb}
\usepackage{epsfig}
\usepackage{amsmath,amssymb,amsthm,mathrsfs,amsfonts,dsfont}
\usepackage{adjustbox,lipsum}
\usepackage{amsfonts}
\usepackage{epsfig}
\usepackage{amssymb}
\usepackage{amsmath}
\usepackage{amsthm}
\usepackage{subfigure}
\usepackage{multirow}
\usepackage{rotating}
\usepackage{graphicx}
\usepackage{tabularx}
\usepackage{array}
\usepackage{anyfontsize}
\usepackage{color,soul}
\usepackage{graphicx,dblfloatfix}
\usepackage{epstopdf}
\usepackage{blindtext}
\usepackage{amsmath}
\usepackage{amsthm,amssymb,amsmath,bm}
\usepackage{subfigure}
\usepackage{amsfonts}
\usepackage{epsfig}
\usepackage{amssymb}
\usepackage{amsfonts}
\usepackage{amsmath}
\usepackage{cite}
\usepackage{hyperref}
\usepackage{graphicx}
\usepackage{fancyhdr}
\usepackage{subfigure}
\usepackage[subfigure]{tocloft}
\usepackage[font={small}]{caption}
\usepackage{subfigure}
\usepackage{tabularx}
\usepackage{tcolorbox}
\usepackage{cite}
\usepackage[linesnumbered,ruled,vlined]{algorithm2e}
\SetKwInput{KwInput}{Input}
\SetKwInput{KwOutput}{Output}
\usepackage{amsthm,amssymb,amsmath,bm}
\usepackage{graphicx}
\usepackage{fancyhdr}
\usepackage{subfigure}
\usepackage[subfigure]{tocloft}
\usepackage[font={small}]{caption}
\usepackage{subfigure}
\usepackage{tabularx}
\usepackage{cite}
\allowdisplaybreaks

\newtheorem{Assum}{Assumption}
\usepackage{epstopdf}

\begin{document}
	\title{ \huge 
		Proactive and AoI-aware Failure Recovery for Stateful NFV-enabled Zero-Touch  6G Networks: Model-Free DRL Approach
	}
				\author{\IEEEauthorblockN{Amirhossein Shaghaghi, Abolfazl Zakeri, \IEEEmembership{Student Member, IEEE}, Nader Mokari, \IEEEmembership{Senior Member, IEEE}, Mohammad Reza Javan, \IEEEmembership{Senior Member,
					IEEE}, Mohammad Behdadfar and Eduard A Jorswieck, \IEEEmembership{Fellow,
					IEEE}
			\IEEEcompsocitemizethanks{\IEEEcompsocthanksitem A. Shaghaghi and M. Behdadfar are with the School of engineering,
				IRIB University, Tehran, Iran (email: behdadfar@iribu.ac.ir). 
				A. Zakeri and
N. Mokari are with the Department of ECE,
Tarbiat Modares University, Tehran, Iran (email: \{abolfazl.zakeri and  nader.mokari\}@modares.ac.ir). 
				Mohammad R. Javan is with the Department of Electrical and Robotics Engineering, Shahrood University of Technology,
				Shahrood, Iran (javan@shahroodut.ac.ir).
				Eduard A. Jorswieck is with  TU Braunschweig, Department of Information Theory and Communication Systems, Braunschweig, Germany (jorswieck@ifn.ing.tu-bs.de).
		}}
		\thanks{This work was supported by the joint Iran national science foundation (INSF) and German research foundation (DFG) under grant No. 96007867.}}
	\IEEEdisplaynontitleabstractindextext
	\IEEEpeerreviewmaketitle
\vspace{-4em}
	\maketitle
	\begin{abstract}
		In this paper, we propose a Zero-Touch, deep reinforcement learning (DRL)-based Proactive Failure Recovery framework called ZT-PFR for stateful network function virtualization (NFV)-enabled networks. To this end, we formulate a resource-efficient optimization problem minimizing the network cost function including resource cost and wrong decision penalty. As a solution, we propose state-of-the-art DRL-based methods such as soft-actor-critic (SAC) and proximal-policy-optimization (PPO). In addition, to train and test our DRL agents, we propose a novel impending-failure model. Moreover, to keep network status information at an acceptable freshness level for appropriate decision-making, we apply the concept of age of information to strike a balance between the event and scheduling-based monitoring. Several key systems and DRL algorithm design insights for ZT-PFR are drawn from our analysis and simulation results. For example, we use a hybrid neural network, consisting long short-term memory layers in the DRL agent's structure, to capture impending-failure's time dependency.
	\end{abstract}

\vspace{-1em}
\section{introduction} \label{Intro}
\subsection{Motivation and State of The Art}
Nowadays, with the exponential growth of data traffic and new emerging services with ultra-responsive real-time network connectivity and high-reliability requirements such as remote healthcare, self-driving cars, and industrial automation, consistency, and reliability of a network become more important than ever \cite{NFV-iot}. Fulfilling these service requirements in an efficient and flexible manner is challenging. To this end, the next generation of wireless
			communication called sixth-generation (6G), with the 
			support of artificial intelligence, ultra-reliability, and zero-touch network management, is expected to emerge in near future \cite{6gdir,6gartificial}.
To tackle this challenge, network function virtualization (NFV) and software defined network (SDN) have emerged as promising technologies to provide flexible and scalable network and efficient resource management \cite{informatics-nfv}. NFV decouples network functions (NFs) from the proprietary hardware, which allows service providers to run virtualized NFs (VNFs) with different functionalities on top of a common physical node as software. 
Based on the desired services, a tenant requests a set of network services in the form of a service function chain (SFC). SFC is a sequence of VNFs fulfilling end-to-end (E2E) service demands in a specific order.
Packets processed in each VNF are steered to other VNF in the sequence for further processing until the last VNF \cite{flow_migration}. 
\\\indent 
Despite flexibility and resource efficiency achieved by network softwarization, it poses new concerns especially in terms of reliability and consistency of services \cite{Fault_Elsi}. VNFs are software running on physical nodes, which are vulnerable to various faults and problems such as physical node failure and software malfunctions \cite{ETSI_fault}. To encounter failure problems and enhance network reliability and performance, deploying backup instances is indispensable \cite{Fault_Elsi}.
Many VNFs are state-dependent and states are updated according to the traffic traversing through them, for example, a virtual network address translation (NAT) updates its states based on IP and MAC addresses of the new connected devices \cite{statelet}. 
If a failure happens in a stateless VNF, software defined network (SDN) controller will simply reconfigure the flow path through a deployed backup instance. But for a stateful VNF to maintain robustness and consistency of SFC, backup VNF's state must be synchronized with the active VNF\footnote{For example, if a NAT fails, backup VNF instance must receive the most recent updates which were made by the active VNFs, to guarantee seamless recovery procedure.}.
Therefore, state synchronization for seamless failure recovery in stateful VNFs is necessary and challenging. In this paper, we focus on the case where all VNFs are stateful. 
\\\indent
Generally,  two schemes for the failure recovery exist which are called \textbf{proactive failure recovery} (PFR) and \textbf{reactive failure recovery} (RFR) \cite{proactive-recovery,proactivemag}. 
Failure recovery is a  procedure that consists of three main stages as 1) launching backup VNF and image migration,  2) flow reconfiguration, and 3) state synchronization. Executing each stage imposes a considerable delay resulting in not only network performance degradation but also service level agreement (SLA) violation due to high service interruption time.
By failure prediction, PFR method can decrease recovery delay by engaging some stages of the failure recovery procedure before the failure manifest. For example, PFR can save flow rescheduling and backup lunch delay, by initiating these stages beforehand \cite{proactivemag}. At this point, if we manage to recover failed VNF in  a PFR manner, the network performance could be greatly enhanced by reducing the failure recovery interruption time.  
This motivates us to propose a PFR framework for future softwarized networks.
 \\\indent
 At the same time, a fully automated and self-managed network is a new paradigm for future networks, e.g, six-generation (6G)\footnote{Recently,  fifth generation of wireless networks is deployed and its evolution towards 6G has been started  \cite{6g1}.}, which can be realized by machine learning (ML) and softwarization \cite{AI_Mag, aitarik}.
 	Recently, deep reinforcement learning (DRL), as an important branch of ML, has made a significant breakthrough and achieved superhuman results, even without human knowledge in strategic games \cite{alpha_zero, Dota2_openai,sutton2018reinforcement}. Also, DRL has achieved good results in the context of NFV such as SFC embedding \cite{jsac_placement,else_datadriven,IOT_embedding_twc}. 
 	 An SFC-driven network could include plenty of physical and virtual entities, and the dynamic changes in each entity's status would cause a high-dimensional and complex state space. Therefore to mitigate the failure consequences in high-dimensional state space, a variety of actions would be possible.
 	  \\\indent To ensure demanded network reliability and robustness in emerging network technologies, immediate reactions for the dynamic changes and events in the networks are necessary. Because of the mentioned challenges, it is difficult for a human orchestrator to predict the likelihood of failures, based on the received information, and to take simultaneous and optimal actions in the network. Benefiting from deep neural networks, DRL is capable of handling high-dimensional state-action spaces and  automatic reactions for the changes in network status. By exploring the underlying environment, e.g., NFV-based network, and evaluating the network status, based on the  monitored information, DRL can effectively evaluate the underlying physical/virtual network entities. By experience gained from the exploration, DRL can learn to adjust and conduct better actions in each situation.
 	Therefore,  we expect DRL to enlighten the solution of our PFR framework. Moreover, modeling the network dynamic is difficult and maybe impossible in practical cases. Therefore, we tend to use model-free DRL, in order to learn network dynamics by training on sample-based experience to make zero-touch automatic decisions. 
 	Besides, to realize PFR, the DRL agent\footnote{In our network the DRL agent is the network orchestrator.}, should access all relevant and necessary information of underlying physical/virtual network to make appropriate decisions in each state \cite{AoI_SDN}.   Therefore, the freshness of this information becomes crucial.  
 We apply
 	age of information (AoI) concept to quantify the freshness \cite{AoI_Network_Metric}
 	via  introducing maximum  AoI as a tolerable freshness constraint.
 	\\\indent 
 	 By combining our \textbf{PFR} framework and the \textbf{model-free DRL} method, we propose a novel intelligent PFR for stateful VNFs, which is called \textbf{Zero-Touch PFR} (ZT-PFR). \footnote{The code for reproducing our results is available at \url{https://github.com/wildsky95/ZT-PFR}.}

\vspace{-1.3em}
\subsection{Main Contributions and Research Outcomes}
In this paper, we propose a novel ZT-PFR scheme to maximize the stateful SFC reliability and ensure network service consistency. Considering resource limitations and maximum tolerable service interruption time caused by failure, our aim is to maintain a highly reliable and resource-efficient network. We devise state-of-the-art soft actor-critic (SAC) \cite{SAC} and proximal policy optimization (PPO) \cite{ppo} model-free DRL methods to automate and optimize our proposed framework. Moreover, to construct an environment simulator to train and test our proposed DRL-based framework, we propose a simulated model of impending-failure in NFV-based networks. Besides, to capture the time dependent features, we equip the agents with long short-term memory (LSTM) layers. Additionally, To provide the DRL agent with the needed information for appropriate decision making, we model an event triggered and scheduling-based AoI-aware monitoring scheme, to observe the network status and guarantee the necessary freshness of information.

 The main contributions of this paper are listed as:
\begin{itemize}
	\item  Considering the dynamics of NFV-enabled networks, we model a PFR framework for embedded stateful SFCs. To this end, we consider a 3-stage failure recovery procedure   aiming to manage resource efficiency and to minimize SLA violations caused  by service interruptions. Moreover, we formulate the PFR as an optimization model aiming to minimize a weighted cost including resource cost and wrong decision penalty.
	\item To realize the proposed ZT-PFR, we adopt state-of-the-art model-free agents, i.e., PPO\cite{ppo}  and SAC\cite{SAC}, and customize them for our model. In addition,  we use a hybrid neural network (NN) consisting of long short-term memory (LSTM) layers.
	Accordingly, in order to train and test our agents, we  design a novel simulated network environment considering the impending-failure concept.
	\item We propose an AoI-aware event-triggered and scheduling-based monitoring scheme, to provide the necessary information freshly to decision-maker (controller), based on the network dynamic.
\item
	  	Several simulation scenarios are provided to assess our ZT-PFR algorithm.  Several key DRL algorithm design insights are drawn from our analysis and simulation results. For example, we use LSTM layers in the DRL agent's NN structure to capture impending-failure's time dependent features. Also, we evaluate the discount-factor influence on sequenced decision making \cite{sutton2018reinforcement} for ZT-PFR.  
	  	Our model shows promising performance in resource efficiency and ZT-PFR in a fair comparison with baselines.

\end{itemize}
\vspace{-1.5em}
\subsection{{Paper Organization}}
	The rest of this paper is organized as follows. The related works are discussed in Section \ref{Related works}. System mode and problem formulation are stated, respectively,  in Section \ref{SystemModel} and Section \ref{Problemformuation}.  Sections \ref{solution} presents the proposed solution. Finally, simulation level evaluation and conclusion remarks are expressed in Sections  \ref{simulation} and \ref{conclusion}, respectively.
\vspace{-1em}
\section{Related Works}\label{Related works}
The proposed ZT-PFR is built upon backup placement, SFC flow reconfiguration, and failure recovery procedure. In this section, we review recent studies on these topics.
There are a few studies on backup placement and recovery  in recent years, and  most of them focus on stateless backup placement and availability optimization\cite{bu_shareconf,multi-tenancy,bu_resource_allocation,ghazizadehreliability}. For example, the authors of \cite{bu_shareconf} propose a stateless backup provisioning scheme that starts by deciding on the number of shared backups and their placements. To improve the resource utilization efficiency, the authors of \cite{multi-tenancy} introduces a new sharing mechanism of redundancy and multi-tenancy technology. \cite{bu_resource_allocation} proposes a backup resource allocation model for middleboxes  considering the importance of functions and both failure probabilities of functions and backup servers. \cite{ghazizadehreliability} studies a reliability-aware resource allocation algorithm using the shared protection scheme with active-standby redundancy for SFC. Considering stateful VNFs, \cite{Fault_Elsi}  studies optimization problems on fault-tolerant stateful VNF placement in cloud networks. The authors consider the VNFs and backup resources demand of incoming SFC request as a constraint to optimize the deployment problem. The author takes VNF state synchronization into consideration, however, VNF state update bandwidth demands and SLA violations due to service interruption are not investigated. Moreover, in \cite{flow_migration}, the authors study a seamless stateful flow reconfiguration and state synchronization problem considering the interruption time and bandwidth limits.
\\\indent
The aforementioned studies do not consider the failure prediction approach to solving the recovery problem. However, there are studies on the failure prediction with data-driven ML methods in NFV and cloud-based networks\cite{fault-cloud,proactive-recovery,impending_Fault,huangproactive}. To the best of our knowledge, only \cite{proactive-recovery} and \cite{huangproactive} take the failure prediction into account and study a proactive path restoration strategy in the NFV-based networks. The mentioned studies propose a master-slave VNF structure to ensure service consistency and consider that each active VNF (master) is supported by some backup VNF instances (slaves). To mitigate the interruption delay, \cite{huangproactive} proposes launching virtual machines (VMs) and flow reconfiguration before failure manifests and  migrating the real-time master VNF's image to a successive slave, afterwards. In best-case scenario, the SFC would be interrupted during image migration. In contrast to \cite{proactive-recovery} and \cite{huangproactive}, to reduce the interruption delay even more and to take service SLA into account, we propose a detailed image migration process based on \cite{statelet,flow_migration}, and break this process into two stages namely snapshot migration and statelet\footnote{\textcolor{black}{Statelets are compact representations of information in incoming packets that change the state of a VNF after snapshot migration\cite{statelet}.}} synchronization. In order to limit the interruption time to a manageable low statelet synchronization delay, the snapshot migration should be done before failure manifest. Furthermore, we take resource efficiency and DRL-based automation into account.
\\\indent 
As concluded from the aforementioned related works and to the best of our knowledge, there is no work on the DRL-based ZT-PFR in softwarized high-reliability  future networks. 
\vspace{-1em}

\section{System Model and Failure Recovery Preliminary }\label{SystemModel}
\vspace{-0.5em}
In this section, we describe the considered system model for the proposed proactive recovery procedure.
First of all, we present the main symbol notations as follows.
Vector and matrix
variables are indicated by bold lower-case and upper-case letters,
respectively.  $|\cdot|$ indicates the absolute value, and $ \Bbb{N}_{+} $ indicates the positive integer values. 
$\Bbb{E}\{.\}$  denotes the statistical expectation, and $\oplus $ denotes the logical XOR function.
\vspace{-1.2em}
\subsection{Physical Network}
	
\begin{figure}[!ht]
 	\centering
 	\includegraphics[width=0.95\linewidth]{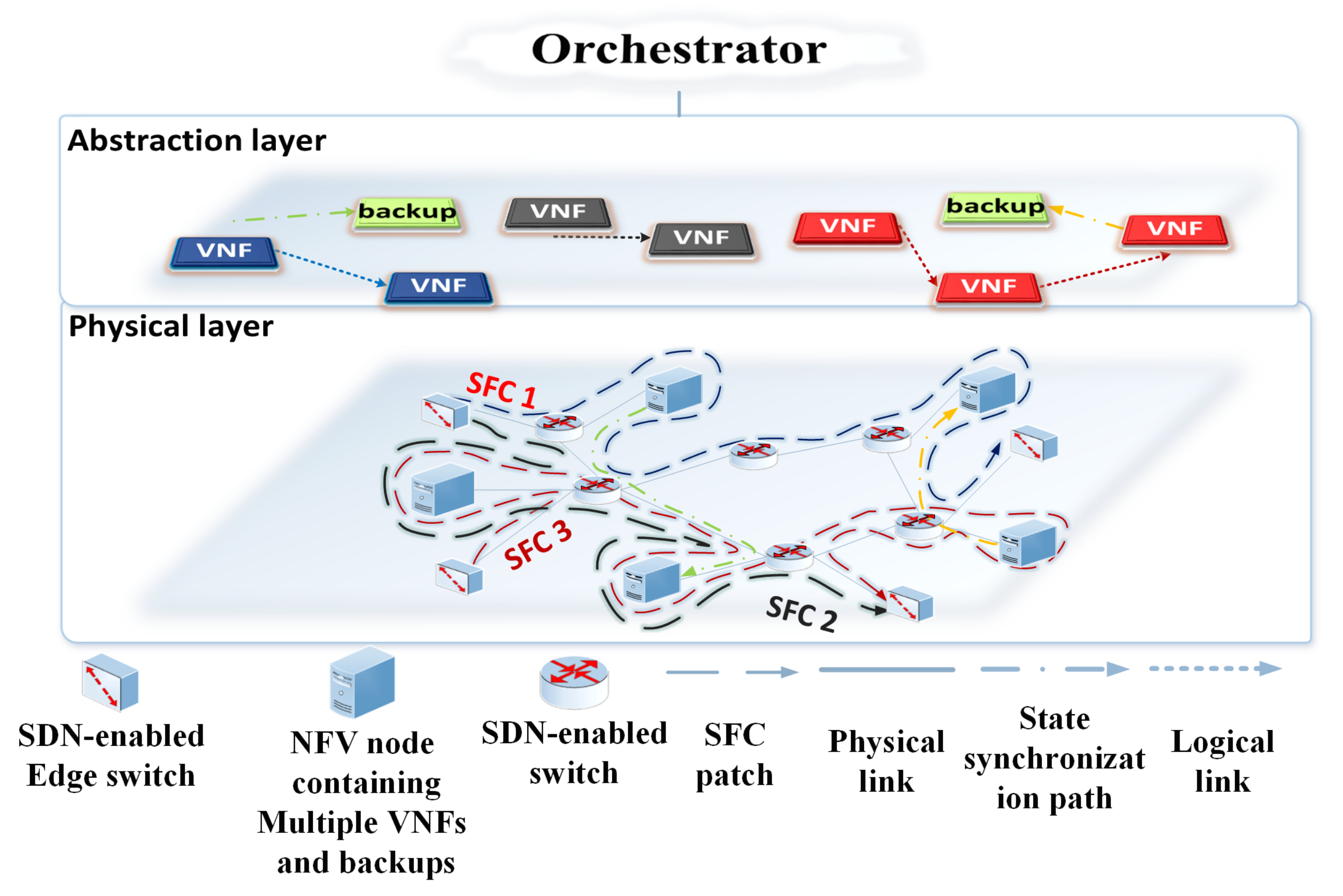}
 	\caption{ 
 		An example of the considered network structure, illustrating the physical layer and virtual abstraction of the embedded SFCs, and their backups and synchronization links.}
 	\label{System model}
\end{figure}
	As depicted in Fig. \ref{System model}, the considered physical network is presented as graph $ \mathcal{G}_P=\big(\mathcal{N},\mathcal{L}\big)$, where $\mathcal{N}$ and $\mathcal{L}$ represent the sets of all physical nodes and links, respectively. Furthermore, $ m,n \in \mathcal{N}$ represent two different nodes and $l_{mn} \in \mathcal{L}$ represents the physical link connecting nodes $m$ and $n$.
	In the network, $\mathcal{N}$ consists of NFV-nodes and SDN-enabled forwarding devices, where all are orchestrated and managed by a centralized orchestrator. The NFV nodes provide processing resources for VNFs, and switches, i.e., forwarding devices, forward traffic from incoming links to outgoing links.  The main parameters are listed in Table \ref{Table_Not}. Note that the parameters superscripted with a prime symbol are related to backup VNF properties. Also, we assume that the continuous-time is slotted into  positive numbers indexed by $ t\in\Bbb{N}_{+} $.
		\begin{table}
		\caption{Table of the main notations/parameters and variables}
		\label{Table_Not}
		\centering
		\renewcommand{\arraystretch}{.9}
		\begin{tabular}{c||l}
			\hline\textbf{Notation(s)} & \textbf{Definition} \\
			\hline  \multicolumn{2}{l}{~~~~~~~~~~~~\textbf{Notations/parameters}}
			\\
			\hline $N/\mathcal{N}/n$ & Number/set/index of physical nodes \\
			\hline $L/\mathcal{L}/l_{mn}$ & Number/set of physical links/index of physical \\& link   connecting physical node $m$ and $n$ \\
			\hline$P / \mathcal{P}/p$ & Number/set/type of resources \\
			\hline$C_n^p / C_{mn}^{BW} $ &  maximum customizable  amount of,  resource\\& type $p$  in node $n$/physical link $l_{mn}$ bandwidth \\
			
			\hline$W_n^p(t) / W_{mn}^{BW}(t) $ & Allocated ratio of, resource type $ p $ in node $n$/\\& physical link $l_{mn}$ bandwidth, at time slot $t$ \\
			\hline$ t/\delta$ & Index/duration of each time slot 
			\\	\hline	$ K/\mathcal{K}/k $&  Number/set/index    of embedded SFCs
			\\
			\hline $\mathcal{H}_k/H_k$ &The set/number of sequenced VNFs in SFC $k$ \\
			\hline $V_h^k$ & The $h$-th VNF in  SFC $k$  \\
			\hline $\mathcal{\varDelta}_k/\mathcal{\sigma}_k$ &Maximum down time/traffic rate (packet/s) of\\&  SFC $k$  \\
			\hline $ \phi_{(k,h)}^p /U_h^k$ & The amount of resources type $p$ needed/the\\&  resource use cost of a backup instance for $V_h^k$ \\
			\hline $ \alpha^k_h(t)$ &  Ratio coefficient managing the  backup\\&  placement cost influence for  $V_h^k$ on each state\\
			\hline $Z_h^k(t)/d_h^k/ b^k_h(t)$  & Accumulated statelet size (in bits)/delay/\\&bandwidth of logical statelet synchronization\\& link of $V_h^k$  \\
			\hline $P_{\textit{nn}}/P_{\textit{nw}}$ & State transition probability from normal to \\& normal/warning\\
			\hline $P_{\textit{ww}}/P_{\textit{wc}}/P_{\textit{wc}}$& State transition probability from warning to\\& warning/critical/normal\\
			\hline $q_v$& Number of least time slots VNF $v$ would \\& stay in   warning state\\
			\hline $\theta_v(t)$& State information AoI of  VNF $v$ at time $ t $\\
			\hline $\kappa_v^s(t)$& AoI constraint depending on VNF $v$ and its \\& state $s$  at time slot $t$\\
			\hline ${\rho}_h^k(t)$& Binary variable indicating weather if $v_h^k$ is in\\&  critical state at time slot $t$\\
			\hline \multicolumn{2}{l}{~~~~~~~~~~~~\textbf{ Optimization Variables}} 
			\\
			\hline$y^{\prime (k,h)}_n (t)$ & Binary variable for embedding VNF $ (k,h) $ in\\&  physical node $ n $\\
			\hline$y^{\prime (k,h)}_{mn} (t)$ & Binary variable for embedding VNF $ (k,h) $ in\\&  physical link $ l_{nm} $
			\\ \hline $m_h^k(t)$ &Binary variable indicating if $ V_h^k $ is supported \\&  by a backup at time slot $ t $
			\\
			\hline
			${\beta}_h^k(t)$& Binary variable for failure recovery decision \\& on  $ V_h^k $  at time slot $ t $
			\\\hline
		\end{tabular}
	\end{table}

	We consider each NFV-node $ n $ provides $P$ types of resources indicated by set $\mathcal{P}$, where $p \in \mathcal{P} = \{1,\dots,P\}$ defines the types of resource, e.g., CPU, memory, and storage. We use $C_n^p$	to represent the maximum customizable  amount of  resource type $p$, in  each NFV-node $n$. Also, $C_{mn}^{BW}$ denotes the bandwidth capacity of physical link $l_{mn}$. We use $W_n^p(t) \in [0,1]$ and $W_{mn}^{BW}(t) \in [0,1]$, to indicate the available ratios of resource type $p$ in node $n$ (available portion of $C_n^p$) and available bandwidth in link  $l_{mn}$, at each time slot $t$, respectively. 	The SFCs properties are characterized in the following.
	 \begin{Assum}
	 	\label{assum1}
		We assume that the SFC embedding problem is previously solved and all requirements such as average delay are ensured. Therefore, the SFC embedding problem is not the focus of this paper. VNF embedding optimization can be done similar to  VNF embedding methods proposed in \cite{else_datadriven,flow_migration,jsac_placement,IOT_embedding_twc}.
	\end{Assum}
\vspace{-2em}
\subsection{ Embedded Services Properties}
Let $ \mathcal{K}=\{1,\dots,K\} $ be the set of $K$ embedded SFCs in the network, indexed by $ k $. Each SFC has a specific VNF sequence and SLA requirements indicated by a tuple as follows:
\begin{align}
\text{SFC}_k=\Big(\mathcal{H}_k,\mathcal{\varDelta}_k,\mathcal{\sigma}_k \Big),~\forall k\in\mathcal{K},
\end{align}	
where $\mathcal{H}_k=\{1_k,\dots,h_k,\dots,H_k\}  $ denotes the set of sequenced VNFs in SFC $k$ and 
 $\mathcal{\varDelta}_k \in \Bbb{N}_{+}$ is the maximum tolerable down time\footnote{Maximum service interruption time, which is not recognizable for users, i.e., SFC's maximum tolerable interruption time \cite{flow_migration}.}. Moreover, $\mathcal{\sigma}_k $ denotes the traffic traversing SFC $k$ in packets per second.
   We define the set $\mathcal{V}$ consisting of all embedded VNFs in the network and $\mathcal{E}$ as the set of all logical links between each two logically connected VNFs.  Also, we use  $V_h^k \in \mathcal{V}$ to denote the $h$-th VNF in  SFC $k$. Besides,
    each VNF's reliability could be enhanced by backup provisioning \cite{multi-tenancy}.    
    \\\indent
    It is worthwhile to mention that the reliability of a service highly depends on the reliability of underlying SFCs. At the same time, the reliability of each SFC is obtained from the dependant  VNFs.  
    Therefore, failure and fault occurrence would result in VNF service quality degradation and SFC SLA violation. 
    As discussed, our effort in this paper is to design a proactive failure recovery. 
    Next, we discuss our failure model. 
 \vspace{-2em}
\subsection{Failure Model}\label{Failure_model}
Following \cite{ETSI_fault}, failures can occur in VNFs and physical nodes due to numerous  reasons such as natural disasters in the location of physical servers, software malfunctions, CPU overload, and temperature threshold violation. 
Typically, when a failure occurs in a SFC, users will automatically retry to continue the connection, and if the orchestrator could mitigate the failure impact before a recognizable time interval\footnote{In this paper, we refer to the recognizable time interval as maximum tolerable  time.}, users would not experience the service interruption caused by failures\cite{ETSI_fault}. This is the point where we try to minimize the interruption time as one of the main results of PFR\cite{huangproactive, proactivemag}.  
Most failures could be forecast by monitoring status information (e.g, resource overload and temperature) of VNFs. In this paper, these types of failures are denoted as impending-failures \cite{impending_Fault}. Accordingly, we propose a model to simulate  the notion of the impending-failure in a NFV-enabled network as follows:
\subsubsection{VNF States and State  Transition Model}
\begin{figure}[!ht]
	\centering
	\includegraphics[width=0.95\linewidth]{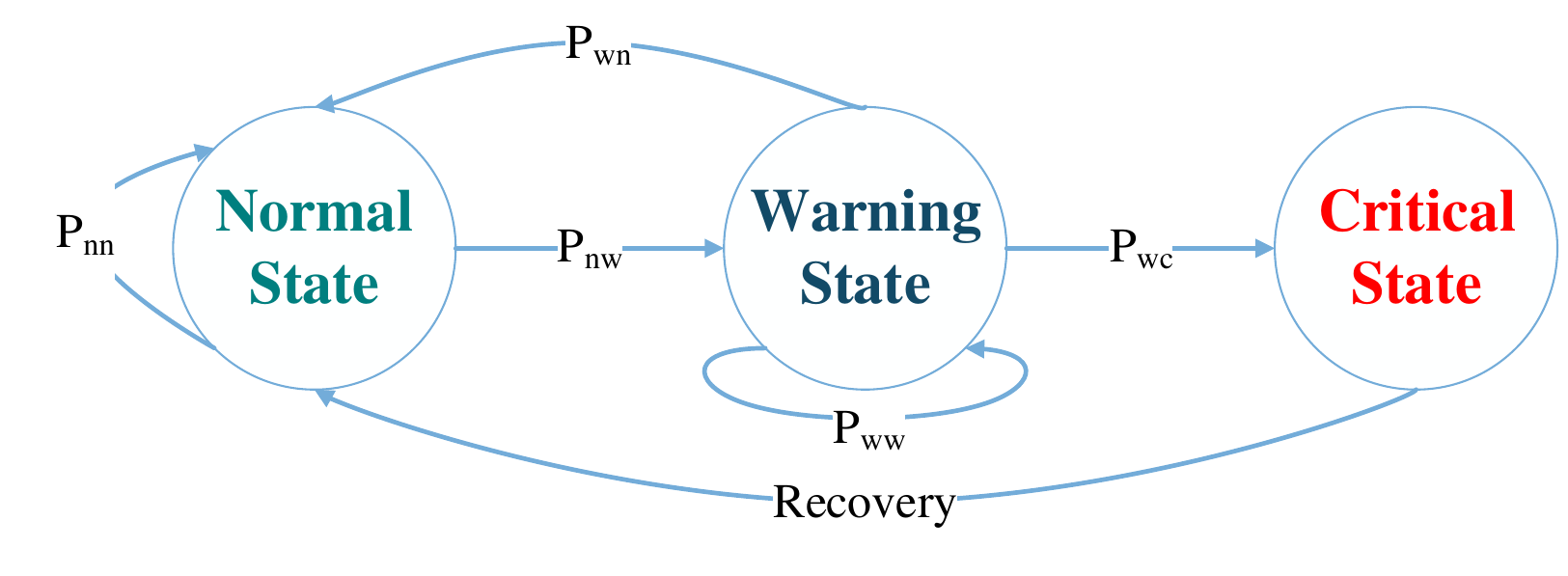}
	\caption{The VNF state transition model at each time slot $ t $.}
	
	\label{State_Tra}
\end{figure}

	An impending-failure in NFV-based networks could be predicted by ML approaches using network infrastructure information, and observing event severity and  service degradation patterns\cite{ETSI_fault,huangproactive,impending_Fault,proactive-recovery}. Our aim in this paper is to prepare the DRL-agent to deal with impending-failures proactively.   
\\$\bullet$ \textbf{VNF States Model:}
The model is inspired by ITU standard X.733\footnote{The ITU.X733 standard specifies a systems management level alarm reporting function, which classifies event severity in a system. There are studies on intelligent event severity detection in the NFV literature\cite{gupta2017fault, shao2018highly}.} \cite{ITU_fault} and  ETSI NFV; Resiliency Requirements\footnote{ETSI has published collection of documents about standardization of NFV such as \cite{ETSI_fault, ETSI_fault1, ETSI_fault2}. Each document standardizes different aspects of NFV applications. For more information visit \url{https://www.etsi.org}} \cite{ETSI_fault} where dormant fault, active fault, and fault management in NFV are defined. Also, four levels of severity of alarms have been defined in ITU standard X.733: Critical, Major, Minor, and Warning. The critical alarm appears when the service can no longer be provided to the user. Major alarm indicates the service affective condition while minor means no current service degradation is there, but if not corrected may develop into a major fault. A warning is an impending service affecting fault or performance issue.
	In the defined  model, each VNF could be in one of three defined states namely \textbf{normal, warning, and critical}. The current state of each VNF depends on the occurred events severity, e.g., overload severity and temperature threshold violation severity. Accordingly, based on the state of each VNF, the orchestrator should make the corresponding decision on the suitable action for each VNF, simultaneously. The properties of the  mentioned states are defined as follows:
	In the \textbf{normal state}, the VNF works normally, event severity is at a tolerable level and service is not degraded. In the \textbf{warning state}, some technical and physical events cause VNF service degradation, and the service needs some maintenance efforts to prepare for a possible failure. Finally, in the \textbf{critical state}, the event severity reaches a crucial level that we assume that the VNF would fail during the time slot, and immediate recovery action is necessary. 
	To simulate the impending-failure concept, we assume each VNF state transition follows the model illustrated in Fig. \ref{State_Tra}. The transition probabilities in each time slot illustrate each VNF's next state likelihood. \footnote{We consider three states for each VNF by merging the ITU.X733 alarm reporting concept and ETSI NFV-Resiliency service degradation.}
	\\$\bullet~$\textbf{State Transition Model:} 
	As characterized in Fig. \ref{State_Tra} and Table \ref{state-tran},
	following our proposed model, if the VNF's state is normal at the beginning of time slot $t$, the VNF continues in the normal state during the mentioned time slot with probability of $P_{\textit{nn}}$, or its state changes to warning with probability of $P_{\textit{nw}} = 1-P_{\textit{nn}}$. Note that, to simulate impending-failure in our model, we assume that, if a normal state turns to a warning state during a time slot, the VNF  will stay in the warning state for at least $q_v$ time slots after the incident \footnote{The reason for this assumption is to simulate the impending-failure concept and possible fault and error correction time \cite{ETSI_fault,ITU_fault}}. Moreover, if the VNF is in warning state at the beginning of time slot $t$ and $q_v$ time slots are passed, the VNF continues in warning state with probability $P_{\textit{ww}}$, or the state turns to critical with probability $P_{\textit{wc}}$, or turns back to the normal state with probability $P_{\textit{wn}}$.
	Because of continuous service degradation, i.e., impending-failure \cite{ETSI_fault}, we consider that, the more time slots  VNF stays in warning state, consequently it would be more probable to turn to  critical state. Therefore we assume  $P_{\textit{wc}}$ will grow by  $P_{\textit{wc}}\times (\text{number of steps in warning state} - q_v) \leq 1$ \footnote{This factor is defined to simulate the possibility of continuous service degradation leading to a failure \cite{ETSI_fault,ITU_fault}.}. Notably, $P_{\textit{ww}}, P_{\textit{wc}}, \text{ and } P_{\textit{wn}}$ must add up to $1$. Finally, if a VNF's state changes to critical, it stays there unless the recovery procedure is completed. After recovery procedure, the VNF's state turns back to normal, and continues its service.
\subsubsection{Monitoring and State Freshness} \label{monitoring}
	 According to our DRL approach to PFR, the orchestrator must be provided with all the needed information
	 for decision making \cite{sutton2018reinforcement}.
	  We assume an event-triggered and dynamic scheduling based monitoring scheme  \cite{AoI_Infocom,AoI_Mon}. If a transition to the new state occurs in VNF, or its scheduling time arrives, the VNF transmits its own  current state information to the orchestrator.  

	 On the orchestrator side, the information is received with a time stamp. The intention of considering manageable scheduling-based monitoring besides event triggering is: 1) to guarantee the required information freshness on the orchestrator's side and 2) to develop a robust monitoring scheme, in case of data loss and unexpected delays. Manageable monitoring schedule time could improve the efficiency of monitoring resource usage. For example, if a VNF state proceeds to normal, the information freshness is less urgent than in other state types. Therefore, the dedicated resource for intense monitoring could be released.
	 
	 Based on VNF's state, the relevant information on the orchestrator side should be relatively fresh. To take the freshness of these states  information into account, we quantify the information's freshness by the AoI metric. Therefore, the below subsection is dedicated to explaining the proposed  AoI model.

\begin{table}[ht]
	\caption{Descriptions of our VNF state transition model}
	\centering 
	{\begin{tabular}{c|l|l|l} 

		\hline\\ 
		State & Normal & Warning & Critical \\ 
		\hline\\ 
	
		Normal  & $P_\text{nn}$ & $P_\text{nw}$ & 0 \\ &&& sudden failures \\ 
	
		\hline
		\multicolumn{4}{l}{ \textit{the VNF  will stay in the warning state for at least $q_v$ time slots.}}\\
		\hline\\
		Warning & $P_\text{wn}$ & $P_\text{ww}$ & $P_\text{wc}$\\ & Possible  & Continuous & Will grow by the factor \\& fault& service & $P_{\textit{wc}}\times(\text{number of steps in}$ \\ & correction&degradation&  $\text{warning state}-q_v) \leq 1$\\ 
		\hline\\
		Critical & Recovery & 0 & No recovery action\\ [1ex] 
		\hline 
	\end{tabular}}
	\label{state-tran}
\end{table}

\vspace{-1em}
\subsubsection{AoI Model}
As recognized from the name of AoI, it is the difference between the received time and the generation time of the last generated information (packet).
Let $ \theta_v(t)$ denote the AoI of information of VNF $v$ at time $ t $. From the time each information is generated to get to the orchestrator, it suffers a delay (e.g., queueing and propagation) until successfully received. But
we assume the network is delay free\footnote{We assume when the state is monitored at the beginning of time slot, it is received on the orchestrator side with negligible delay \cite{AoI_Info_Mod}.}.

After receiving a state information,  AoI increases with steps as the duration of the time slots. Therefore,  the evolution of AoI is characterized by:


\begin{align}\label{AoI_Evo}
	\theta_v(t) =
	\begin{cases}
		\delta,&\parbox{5cm}{If it is transmitted at the beginning
			 of time slot $ t $}\\
		\\\theta_v(t-1)+\delta& \text{Otherwise}
	\end{cases},
\end{align}
where $ \theta_v(0)= \infty $\footnote{After the latest information update, the AoI value increases by time slot duration, i.e., $\delta$ for each step. Then the orchestrator does not have any information about the  states at time slot $t = 0$. Therefore, the AoI is set to be $\infty$ at the beginning.} and $\delta$ is quantified as the length of each time slot.
\newline
According to the state of each VNF, it is important to optimize the \textit{age} of its information. We consider an AoI constraint in which the AoI should not violate a predefined threshold  in each time slot given by 
	\begin{align}\label{AoI_Con}
		\theta_v(t)\le \kappa_v^s(t),
	\end{align}  
	where $\kappa_v^s(t)$ is defined as constant\footnote{In this paper this parameter is the scheduling time.}. Its value depends on the VNF $v$ and its state $s$ at time slot $t$. For example, if VNF is in warning state, $\kappa_v^s(t) \leq q_v \times\delta $ should be satisfied to guarantee necessary data freshness. Moreover, if VNF is in critical state,  $\kappa_v^s(t)\leq \delta $ should be satisfied. The orchestrator needs to know if its actions mitigated the critical state. Note that AoI optimization, i.e., ensuring network information freshness, is done by tuning the monitor scheduling time.		

As discussed before, 
	 failures can occur at any time,  degrading or interrupting services. Hence, VNF backup provisioning to guarantee service reliability and consistency is indispensable in such networks. In this regard, the failure recovery procedure is described in the following. 
	 \vspace{-1em}
\subsection{Failure Recovery Procedure}
As discussed in Section \ref{Intro}, there exist two schemes for the failure recovery which are called \textbf{proactive} and \textbf{reactive} \cite{proactive-recovery,proactivemag}. For a successful stateful VNF  recovery, some steps are essential. These steps are
discussed in the following. 
\\\indent
The first step is launching a new backup instance for the VNF including allocating the required resources of backup VNF\footnote{In our model, this concept is managed by active-standby method, similar to \cite{Fault_Elsi, proactivemag, huangproactive}} and migrating the latest VNF image (snapshot) \cite{statelet}. The second step is flow reconfiguration, i.e, rescheduling the routing path of backup VNF in the SFC. The final step is statelet synchronization (similar to \cite{statelet}) for the stateful VNF \cite{proactivemag}.  Executing each of the steps imposes a delay which is considerable in the real network and causes network performance degradation and service interruption. 
Obviously, by executing some of the above steps before  failure occurrence, the recovery delay would be significantly reduced. This concept is indicated as \textbf{proactive failure recovery}, which is discussed as follows.  
\\$\bullet$ \textbf{Proactive Failure Recovery:}
In this recovery scheme, the orchestrator could predict the failure in the next time slot. Therefore, it can limit overall recovery delay to synchronization delay by running steps 1 and 2 of the recovery procedure beforehand. 
Note that we could not save synchronization delay, because  every statelet produced until failure must arrive at the backup instance \cite{statelet}. Our second goal is to run the proactive failure recovery procedure at an appropriate time to minimize SFC interruption delay, as explained before.
\\$\bullet~$ \textbf{Reactive Failure Recovery:}
As recognized by the name reactive, in this case, all steps of the recovery procedures (specified before) are executed after  failure occurrence. Therefore, it imposes more recovery delay resulting in high service interruption time and network performance degradation. 

\vspace{-1em}
\subsection{Proposed Proactive Failure Recovery}
\begin{figure}
	\centering
	\includegraphics[width=0.95\linewidth]{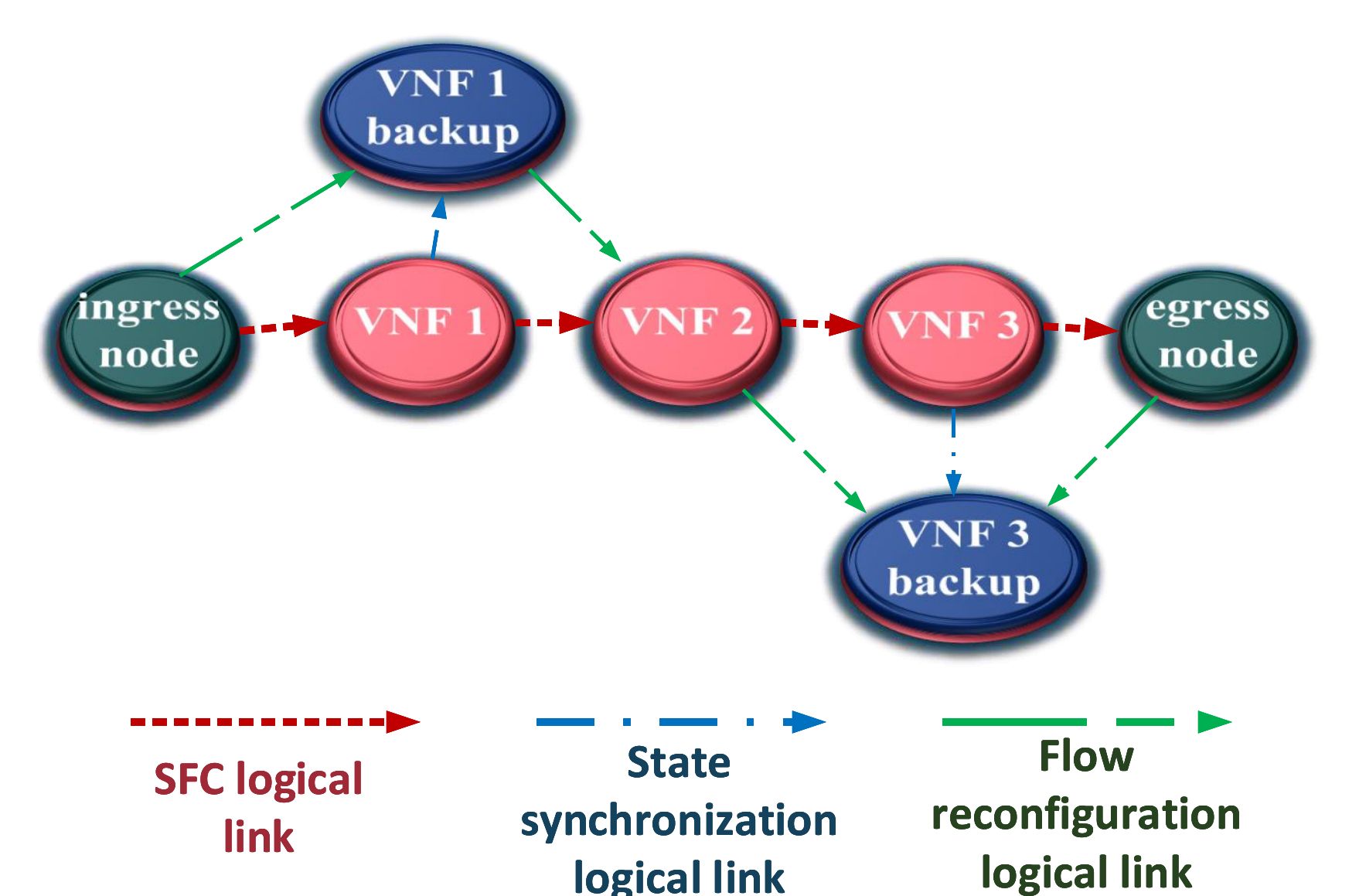}
	\caption{Example of  the considered model for state synchronization and backup recovery procedure by flow reconfiguration, for a single SFC. In this example, the backup placement has been done just for VNF1 and VNF3, and flow reconfiguration links are embedded only when a failure happens.}
	\label{operation}
\end{figure}

	 We consider a dynamic  active-standby failure recovery mechanism where few standby backup instances can be placed and removed in each NFV-node. In case of a VNF failure, the orchestrator transforms the respective backup VNF to an active VNF, and the flow which travels through the failed VNF will be redirected to the new active VNF, i.e., respective backup VNF \cite{proactivemag}. 
	 Each backup utilizes  
	 an amount of resource type $p$ denoted by $\phi_{(k,h)}^p$ for $h$-th VNF in SFC $k$.
	\\\indent
In practical cases,	most VNFs are stateful which means their states update frequently by traversing data. With regards to this, in our case, as seen in Fig. \ref{operation}, we consider that an active VNF's states must be continuously transferred to the backup instance as statelets to provide a seamless flow migration in case of failure \cite{statelet}. 
Moreover, we assume that the statelet update rate of each VNF is linearly proportional to its packet rate $\sigma_k$. 
Accordingly, each backup instance will acquire a logical synchronization link connecting it to the respective active VNF. Due to the limited resources, in the backup placement procedure, each logical synchronization link's bandwidth denoted by $b^k_h(t)$, should take a small predefined amount of bandwidth for VNFs in a non-critical state, denoted by $\phi_{(k,h)}^{BW}$, to maintain the logical synchronization link active, for statelet transfer purposes. Additionally, based on the VNF's type, statelet generation rate and its synchronization sensitivity, the value of the parameter $\phi_{(k,h)}^{BW}$ could be tuned to different values.
\\\indent
We assume that the accumulative statelet size in each time slot is observed by the orchestrator, and  $Z_h^k(t)$ indicates accumulated statelet size of $V_h^k$ in bits till the end of time slot $t-1$, that needs to be transferred in time slot $t$. Hence, the accumulated statelet size in each time slot is the result of constant bandwidth and  packet rate fluctuation. This  leads to a synchronization delay between the active and backup instances. We use $d_h^k(t) = \frac{Z_h^k(t)}{b^k_h(t)}$ to denote the synchronization delay of $V_h^k$ which is caused by $Z_h^k(t)$,  in time slot $t$. This is a dummy delay when VNF works correctly, but in case of failure, this delay must be smaller than the maximum tolerable interruption time $\varDelta_k$ to prevent SLA violation. For example, if $V_h^k$ fails, $b^k_h(t)$ should not be less than $B_{(k,h)}^\text{min}=\frac{Z_h^k(t)}{\varDelta_k}$ to prevent the synchronization delay from exceeding maximum tolerable interruption time \cite{flow_migration}.


\vspace{-1em}
\section{{Problem Formulation}}\label{Problemformuation}
In this section, we formulate the proposed PFR as an optimization problem. 
We assume that based on the orchestrator decisions, the backup instances could be placed and removed for efficient resource utilization purposes. In doing so, we first introduce optimization constraints and then introduce the proposed objective function. 
\vspace{-1em}
\subsection{Network Constraints}
To ensure service consistency, if a VNF enters a critical state, the respective logical synchronization link bandwidth should be optimized to meet the synchronization delay limits. Let $y^{\prime (k,h)}_n (t)$ and $y^{\prime (k,h)}_{mn} (t)$ be binary variables, where $y^{\prime (k,h)}_n (t)$ equals $ 1 $ if $V_h^k$'s backup is embedded in physical node $n$ during time slot $t$, and 0 otherwise. Moreover, $y^{\prime (k,h)}_{mn}(t)$ equals to $ 1 $ if $V_h^k$'s logical synchronization link is embedded in physical link $L_{mn}$ during time slot $t$. It is worth noting that $y^{\prime (k,h)}_n (t)$ equals $ 0 $ for all SDN enabled forwarding devices.  At each time slot, sum of all allocated and released resources should  not exceed the current available resources in NFV-nodes and physical links as:
\begin{multline}
\label{resource_limits}
	\sum_{k \in \mathcal{K}} \sum_{h \in \mathcal{H}_k} (y^{\prime (k,h)}_n(t) -y^{\prime (k,h)}_n(t-1))  \cdot \phi_{(k,h)}^p\\ \leq W_n^p(t) \times C_n^p , \forall p \in \mathcal{P} , \forall n \in \mathcal{N},
\end{multline}
\begin{multline}
\label{bw_limits}
\sum_{k \in \mathcal{K}} \sum_{h \in \mathcal{H}_k} \Big(y^{\prime (k,h)}_{mn} (t) - y^{\prime (k,h)}_{mn} (t-1)\Big) \cdot \phi_{(k,h)}^{bw}\\ \leq W_{mn}^{bw} (t)\times C_{mn}^{bw} , \forall mn \in \mathcal{L}.
\end{multline}
The first parts of \eqref{resource_limits}-\eqref{bw_limits} indicate backup resource allocation and release in each time slot $t$, and the second part indicates the available resource amount in the beginning of time slot. For example, in \eqref{resource_limits}, if a new backup is placed in node $n$ during time slot $t$, $ (y^{\prime (k,h)}_n(t) -y^{\prime (k,h)}_n(t-1))$ would be $+1$, but in case of removing an existing backup and releasing its allocated resources, it would be $-1$. 
\\\indent
We define $m_h^k(t)$ as a binary variable which equals 1, if $V_h^k$ is supported by a backup and recovery steps 1 and 2 are performed, and 0 otherwise. It is given by
\begin{align}
	m_h^k(t) = \mathds{1}\Big(\sum_{n \in \mathcal{N}} y^{\prime (k,h)}_n(t-1) >0 \Big), \forall V_h^k \in \mathcal{V},
\end{align}
where, $\mathds{1}(.)$ is an indicative function, and it equals $ 1 $, if $\sum_{n \in \mathcal{N}} y^{\prime (k,h)}_n(t) >0$, and $ 0 $ otherwise.
In our model, an active VNF and its backup can not be in the same NFV-node, because,  if the respective NFV-node fails, then both backup and active VNF would  fail. This will make backup placement meaningless and the backup placement would be in vain. 
To ensure this matter, we introduce the following constraint:
\begin{align}
\label{eq:6}
\sum_{k \in \mathcal{K}} \sum_{h \in \mathcal{H}_k} y^{\prime (k,h)}_n(t) \cdot y^{(k,h)}_n(t) = 0 , \forall n \in \mathcal{N},
\end{align}
where $y^{(k,h)}_n(t)$ equals 1 if $ V_k^h$ is embedded in NFV-node $n$ in time slot $t$. Also in our model, we consider placing only one backup for each VNF due to resource limitations which is expressed by following constraint:
\begin{align}
\label{eq:7}
\sum_{n \in \mathcal{N}} y^{\prime (k,h)}_n(t) \leq 1 , \forall k \in \mathcal{K} , \forall h \in \mathcal{H}_s.
\end{align}
When a VNF is in a critical state at time slot $t$, the corresponding  synchronization link bandwidth must be reconfigured to prevent synchronization delay threshold violation, i.e., final step (step 3) in the recovery procedure, which is formulated as:
\begin{align}
\label{synch_delay}
	d_h^k(t) \leq {\rho}_h^k(t) \cdot \Delta_k + \big(1- {\rho}_h^k(t)\big ) \cdot \frac{1}{\epsilon},
\end{align}
where ${\rho}_h^k(t)$ equals $ 1 $ if VNF $V_h^k$ is in critical state in time slot $t$, and $ 0 $ otherwise. Also, $\epsilon$ is a small number for ensuring the constraint to be true when the entity works properly.
In the case of a critical state, \eqref{synch_delay} ensures appropriate bandwidth for the logical synchronization link, to synchronize backup and active VNF's state in less than $\Delta_k$.
\\\indent
\vspace{-2em}
\subsection{Objective Function and Problem}
To formulate our objective function, we design a weighted cost function to cover the different aspects of the network cost.
First, in order to optimize and guarantee the backup placement before failure occurrence, we define the first part of our objective function as below:

\begin{align}
\label{cost_a}
	 \Phi_\text{SLA} =\Psi_b \cdot \sum_{k \in \mathcal{K}} \sum_{h \in \mathcal{H}_k} {\rho}_h^k(t) \big(1- m_h^k(t)\big),
\end{align} 
where $\Psi_b$ indicates the imposed cost by service interruption and SLA violations followed by a failure occurrence, which was not supported by a backup, i.e, the VNF's backup was not ready before failure manifest.
\\\indent
Clearly, the backup placement allocates a redundant amount of resources in the network, so it is considered as an overhead to network resource usage. We assume each VNF's backup requires a distinct amount of resource, where this resource usage implicates a resource usage cost denoted by $U_h^k$.
 Therefore, the overall backup placement cost is formulated as follows:
\begin{align}
\label{cost_b}
\Phi_\text{RC}=	\sum_{k \in \mathcal{K}} \sum _{h \in \mathcal{H}} \alpha^k_h(t)   m_h^k(t) U_h^k ,
\end{align}
where $\alpha^k_h(t) $ is a VNF-specific coefficient managing the backup placement cost impact on overall utilization cost. According to the current state and resource requirements of VNF $V^k_h$, the value of $\alpha^k_h(t) $ could be different in each time slot $t$. It is worthwhile to mention that the cost for backup placement in near-critical states should be less than in normal states. This would ensure efficient resource usage based on different states. For example, in the normal states, where the VNF works properly and without any service degradation, the best decision would be to release the allocated resources to the backup VNFs. Therefore, in normal states the cost of the utilized resource for backup is considered to be high.
\\\indent
As mentioned, if the orchestrator detects a critical state  in an entity, it should run failure recovery  procedure. We define ${\beta}_h^k(t)$ equals $1$ if the orchestrator runs the failure recovery procedure, and 0 otherwise.  For the case that the critical detection was wrong, the resource used by the failure recovery procedure would be in vain. Therefore, we assume each wrong critical state detection will cause a penalty cost $\Psi_f$ to the network, which is defined as bellow:
\begin{align}
\label{cost_f}
\Phi_\text{FA}=	\Psi_f \cdot \sum_{k \in \mathcal{K}} \sum_{h \in \mathcal{H}_k} {\rho}_h^k(t) \oplus {\beta}_h^k(t).
\end{align}
In this paper, our objective is to  minimize  the weighted cost via solving the following proposed optimization problem:
\begin{subequations}
	\label{MainP1}
	\begin{align}
\label{Main_Problem_1}
\min_{ \bold{M}, \boldsymbol{\beta},\bold{Y'},\bold{Y}} \quad & \eta_1 \Phi_\text{SLA} + \eta_2 \Phi_\text{RC} + \eta_3 \Phi_\text{FA}\\
\textrm{Subject to} ~~&  
\eqref{resource_limits}- \eqref{synch_delay},
\end{align}
\end{subequations}
where $ ~\bold{M}=[m_h^k(t)],~\boldsymbol{\beta}=[{\beta}_h^k(t)] $, $ \bold{Y}'=[y^{\prime (k,h)}_{mn} (t)] $, $\bold{Y}=[y^{\prime (k,h)}_n(t)]$, and  $ \boldsymbol{\eta}^{T}=[\eta_1~\eta_2~\eta_3]  $ are the fitting parameters. 
\vspace{-1em}
\section{Solution Algorithm}\label{solution}
Problem \eqref{MainP1} is a non-linear integer programming (NLIP) which is generally complicated to solve.
Actually, the optimization variables in \eqref{MainP1} include sequential decisions. Nowadays, it is shown that  DRL has  tremendous performance on the long term sequential decision-making problems without human knowledge \cite{alpha_zero, Dota2_openai}. At the same time, to realize the decisions in an automatic and zero-touch manner, DRL-based solutions are necessary. In addition,  benefiting from deep neural networks, DRL is capable of handling high-dimensional state-action spaces.
These motivate us to propose policy-based model-free DRL solutions,
discussed in the following.
As mentioned before in Sections \ref{SystemModel} and \ref{Problemformuation},
the focus of this paper is to realize PFR considering resource usage efficiency. The embedding variables used in Section \ref{Problemformuation} guarantee the resource limits. As mentioned in Assumption \ref{assum1}, VNF (backup VNF) embedding optimization is not the focus of this paper. Therefore, our actions are related to the PFR steps defined next. 

\vspace{-1.2em}
\subsection{Model-Free DRL and Agents} \label{MDP}
In this paper, we tend to use model-free DRL. 
Policy-based model-free methods directly parameterize the policy $\pi(a|s;\boldsymbol{\theta})$ which is defined as a distribution over  actions $a$ based on  current state $s$ 
and update the neural network parameters $\boldsymbol{\theta}$ by performing gradient ascent on the expected reward.
The expected reward is the reward that an agent receives in a whole episode.
The intention is to create an orchestrator (agent) to learn a PFR  policy, by observing the network  and sampling from the environment. Then, the orchestrator should manage the service reliability and efficient resource usage, in order to minimize the defined cost function. As mentioned in Section \ref{Failure_model}, service degradation or failures can happen any time and on any VNF or NFV node. Therefore, simultaneous actions are needed. Without any knowledge of the environment dynamics, e.g., transition probabilities, the agent starts exploring the environment with a random policy. Since the agent makes sequential decisions along the episodes, it observes the current state $s$ and takes the action $a$ based on its current policy. Then, the environment returns reward $r$ and the new state $s'$ based on the taken action. These trajectories of experience are recorded in the agents experience buffer as the tuple $(s, a, r, s')$. They are used as training data to improve the policy. The states, actions, and reward function in our model are configured as follows.
\\$\bullet~$\textbf{States:} 
  As discussed in Section \ref{Failure_model},
    the $v$'th VNF state, which we feed to the agent, can be in  three types of classes, which is denoted by  $\bold{{S}}^v_{\text{type}}(t)=[s_{\text{N}}^{v}(t), s_{\text{W}}^{v}(t) , s_{\text{C}}^{v}(t)]$, where $s_{\text{N}}^{v}(t)\triangleq1$ indicates the \textit{normal} state, $s_{\text{W}}^{v}(t)\triangleq2$ indicates the \textit{warning} state, $s_{\text{C}}^{v}(t)\triangleq3$ indicates the \textit{critical} state, respectively. 
    Since each VNF could be in one of the states in each time slot, the total state space is $ \bold{S}_{\text{Tot}}(t)= \prod _ {v \in \mathcal{V}}\bold{S}_{\text{type}}^v(t)$.
\\$\bullet~$\textbf{Actions:} 
 We define our actions as  three types for each VNF 
including the backup placement (BP), backup removal (BR), and statelet synchronization (SS). BP includes executing the first and the second steps of the  failure recovery procedure. SS indicates executing the third step of recovery. Finally, the BR action indicates releasing all of the resources allocated by  aforementioned actions.
To realize  PFR, the desired behavior would be to run BP when the state leading to the critical state and executing SS when the critical state is observed. In contrast, in the RFR, after a  critical state is observed all three steps of the recovery have to be executed.  
Also, for resource-efficient PFR, the desired action for the normal state  would be BR. 
\\$\bullet~$\textbf{Reward function:} 
In order to minimize the predefined cost,  i.e., the objective function  \eqref{Main_Problem_1},  the learned policy should take the appropriate actions based on the given state $s$.
 In each time slot $t$, which corresponds to state $s(t)$, the agent samples action $a(t)$ from $\pi(a|s;\boldsymbol{\theta})$, and  receives next state $s(t+1)$ and immediate reward $r_{\text{Tot}}(t)$ from the environment. The agent's goal is to maximize cumulative reward in each episode. Note that minimization of the cost function  could be equivalently converted  to a maximization problem. Accordingly, the first part of the reward function is constructed from the cost function \eqref{Main_Problem_1}  as follows:
\begin{align}
	\label{cst_reward}
	r_1(t) =-\eta \Phi_\text{SLA} - \eta_2 \Phi_\text{RC} - \eta_3 \Phi_\text{FA}, ~ \eta_1,  \eta_2, \eta_3\geq 0,
\end{align}
		where $\Phi_\text{SLA}, \Phi_\text{RC}, \text{and } \Phi_\text{FA}$ are defined by \eqref{cost_a}, \eqref{cost_b}, and \eqref{cost_f}, respectively.
\\\indent
To encourage the agent to take the desired actions, we also add  positive reward $r_2(t)$ to the reward function, including terms as:
  1) positive reward for BR action in the normal state, i.e, $\Phi_\text{BR}$, 2) positive reward for BP action before critical state manifest, i.e, $\Phi_\text{BP}$, 3) positive reward for SS in a critical state, i.e, $\Phi_\text{SS}$, and 4) positive reward for successfully completing the PFR on a failed VNF $\Phi_\text{PFR}$. Accordingly, the $r_2(t)$ is defined as bellow:  
\begin{align}
		r_2(t) = \Phi_\text{BR} + \Phi_\text{BP} +  \Phi_\text{SS} +  \Phi_\text{PFR}.
\end{align}
   Numerical values are specified in Section \ref{simulation}. The additional positive rewards are added to \eqref{cst_reward} to construct the total reward in each step of episode. Accordingly, the total reward is defined as:
   \begin{align}
r_{\text{Tot}}(t) = r_1(t) + r_2(t).
   \end{align}
    
The aim of each DRL agent is to maximize the expected reward through an entire episode of the environment. Note that, maximizing the negative of cost function equals to minimizing it. Accordingly, as the DRL agent converges to a higher reward followed by higher accuracy, our objective function converges to a lower cost.
It can be concluded that maximizing the expected reward by taking correct actions, minimizes the network cost. The numerical configuration is discussed in Section \ref{simulation}. Below, we provide details of the proposed method to find the policy. 
\vspace{-1em}
\subsection{Policy Optimization}
 In the most well known policy optimization method, called \textbf{REINFORCE}, the agent generates data   for a whole episode, based on the current policy. Then, stochastic policy parameters update after each episode by
 \begin{align}
\nabla_\theta\log\pi\big(a(t)|s(t);\theta\big) R(t),
 \end{align}
  where $\nabla_{\theta}$  denotes the gradient with respect to parameter $\theta$, and $\pi\big(a(t)|s(t);\theta\big)$ indicates the stochastic policy, which is parameterized by $\theta$. Therefore, the updates are highly variant, and unstable. It is possible to reduce the variance of this estimate while keeping it unbiased by subtracting a baseline denoted by $b(t)$ \cite{baseline} as:
  \begin{align}
  	\nabla_\theta\log\pi(a(t)|s(t);\theta)[R(t) - b(t)] .
  \end{align}
    A learned estimate of the value function is commonly used as the baseline. Then, the quantity $R(t) - b(t)$, used to scale policy gradient, can be seen as an estimate of the advantage of taking action $a(t)$ in state $s(t)$. Because $R(t)$ is an estimate of $Q_{\pi}(a(t),s(t))$, and $b(t)$ is an estimate of $V_{\pi}(s(t))$, where $Q_{\pi}(a(t),s(t))$ and $V_{\pi}(s(t))$ denote state-action value and state-value function, respectively. The advantage function is defined as:
\begin{align}
\varXi\big(a(t),s(t)\big) = Q\big(a(t),s(t)\big) - V(s(t)).
\end{align}
  This approach is named as actor-critic architecture, where actor chooses action based on policy $\pi$ and $b(t)$ evaluates the action by the value of the state\cite{A3C}.
\\\indent
The \textbf{REINFORCE} method uses the $\log$ probability of the actions, to trace the impact of actions. But there are other functions for this matter \cite{objectivefunc}. Also, we do not want our policy parameter updates to be large in each iteration in on-policy methods, because the agent might get stuck in poor policy and generate data based on that policy. Thus, learning on that data could cause a wrecked policy. To mitigate this problem, \cite{tpro} proposes a Trust Region Policy Optimization, which uses KL-divergence as a constraint or penalty to limit policy parameter updates. But this method has a complex computation and needs lots of processing power. In this paper, we use Proximal Policy Optimization \cite{ppo} with Clipped Surrogate Objective (PPO-CSO), which is a first-order objective algorithm described next. 

\subsubsection{Proximal Policy Optimization (PPO)} 
On the contrary to vanilla policy optimization, where updating parameters for more than one epoch may cause a large policy update, PPO-CSO can train $K$ epochs for each iteration due to limited update of parameters. Thus, PPO-CSO method has better sample efficiency than vanilla policy optimization. 
Let $p(t;\theta)$ denote the probability ratio at time slot $ t $ over $ \theta $ which is defined by
\begin{align}
	 p(t;\theta) = \frac{\pi_{\theta}(a(t)|s(t))}{\pi_{\theta_{old}}(a(t)|s(t))}, ~  p(t;\theta_{old}) = 1, \forall t.
\end{align}
\\\indent
 With limited policy update, we want to maximize the expected reward. Thus, the clipped surrogate objective $L^{\text{CLIP}}(\theta)$  is defined as follows:
\begin{multline}
	L^{\text{CLIP}}(\theta) = \mathbb{\hat{E}}_{t}\{\min[p(t;\theta)\hat{\varXi}(t) ,\\ \text{clip}\big(p(t;\theta), 1- \epsilon , 1+\epsilon\big){\hat{\varXi}}(t)]\},
\end{multline}
where $\epsilon$ is the limiting hyper-parameter. Moreover, a MSE-based objective function for  state-value network at each time slot $ t $ parameterized over $ \theta $ is defined as follows:
\begin{align}
	L^{VF}(t;\theta) = \big\{V^{\text{Targ}}(t) - V_{\theta}(s(t))\big\}^2.
\end{align}
Also, $\Upsilon_{\pi_\theta}(s(t))$ denotes an entropy bonus to ensure sufficient exploration \cite{ppo}. 
 Thus, combining these terms, the aim is to maximize the main objective function defined as:
\begin{multline}
	L _t(\theta) = \mathbb{\hat{E}}_t\{L^{CLIP}(t;\theta) - c_1L^{VF}(t;\theta)\\ + c_2\Upsilon_{\pi_\theta}(s(t))\},
\end{multline}
where $c_1>0$ and $c_2>0$ are influence coefficients.
\subsubsection{Soft Actor-Critic (SAC)}
SAC is an off-policy actor-critic DRL method based on entropy maximization RL framework \cite{SAC}. The algorithm adds an entropy term to the reward function to guarantee sufficient exploration while converging to the optimal solution. 
To adopt DRL-based solution, our agent needs to determine states, actions, and reward value function which are presented in the following \cite{SAC}.
\vspace{-1em}
f
\section{Numerical Evaluation}\label{simulation}
This section presents numerical results to validate and assess our proposed PFR framework and algorithm   under various configurations to compare with baseline. We provide numerical results regarding  different metrics such as desired action accuracy for different parameters.
\vspace{-1.3em}
\subsection{Simulation Setup}
	We consider an NFV-enabled network containing $3$ SFCs, i.e., $K=3$, where each SFC is constructed from $3$ VNFs, i.e., $H_k=3, ~ \forall k$. Therefore, there would be $9$ VNFs, i.e., $V=9$, which are embedded on 5 NFV nodes, i.e., $N=5$, unless otherwise stated. 
Also, we consider that each NFV node provides 3 types of resource, i.e., $P=3$, including: CPU, storage, and memory. Moreover, we assume each BP and SS actions require a random amount of resource and  statelet synchronization bandwidth for each VNF, respectively.
BP and failure recovery require another random amount of resource and  statelet synchronization bandwidth for each VNF, respectively. 
\\\indent
The Python library Networkx is used to simulate our network's topology and structure. A network with fully connected NFV nodes is created by Networkx and SFCs are randomly embedded on top of the physical network. Each VNF working status is presented by the defined state transition model in Fig. \ref{State_Tra} with random transition probabilities, generated at the beginning of an episode. Also, we define that each VNF stays in warning state for at least $q_v=2$ consecutive steps. Therefore we consider $\kappa_v^s(t)=2$ for warning states. Moreover, we defined $\kappa_v^s(t)=1$ for critical states to observe the results of the taken actions. Finally, we consider $\kappa_v^s(t)\ge 2$ for normal states. As mentioned in Section \ref{monitoring}, each VNF transmits its own state if an event occurs or a scheduling time interval arrives.  It is worthwhile to note that the parameter $\kappa_v^s(t)$ denotes the scheduling time. The episode length is $100$ steps. To build the agents, we use a hybrid NN structure made of normal and LSTM layers, which is described in Table \ref{Simulation_Sett}, unless otherwise stated.
\\\indent
To numerically design the first part of the  reward function, i.e, $r_1(t)$, we assume every backup has the same placement cost as $U_h^k = 1$. As discussed in Section \ref{Problemformuation}, to enforce different costs for each state type, the value of $\alpha^k_h(t)$ is considered $1$, $0.1$, and $0$ for normal, warning and critical states, respectively. Moreover, values of $\Psi_b,\Psi_f,\eta_1,\eta_2, \text{and } \eta_3$ are defined to be $1$. Also, as discussed in Section \ref{MDP}, the second part of the reward function, i.e, $r_2(t)$, is designed as follows. $+1$ reward for BR action in the normal state, $+1$ reward for BP action before critical state manifest,  $+1$ reward for executing failure recovery procedure in critical state, and finally $+100$ reward for successfully completing a PFR on a failed VNF. Accordingly, the total reward $r_{\text{Tot}}(t)$ would be constructed by adding up all the mentioned rewards. 
\\\indent
To speed up the training and enhance exploration efficiency, we use a distributed learning method, where multiple agents run in parallel, on multiple instances of the environment.
At each iteration, the PPO agent runs the environment in parallel for 32 times and the SAC agent runs the environment for 16 times in parallel. The generated data trajectories are saved in a buffer as a batched data-set. The agent trains its target policy with generated data for 32 epochs and moves on to the next iteration. For policy evaluation, after every 50 iterations, the agent runs the environment for 50 episodes with the most recent target policy and outputs the evaluation data averaged over episodes. It is worth noting that, to evaluate the policy integrity and robustness, during training, we sample the agent's target policy every 500 iterations, and evaluate it in a new environment with new random parameters.
The results are discussed in the next.
\begin{table}
	\centering
	\caption{
	The NN structure, for example in tuple $(x,y)$, length of the tuple indicates the number of hidden layers, and each entity, e.g., $x$, denotes the number of hidden units or LSTM units.
	}
	\label{Simulation_Sett}
	\begin{tabular}{ c|c}
		\hline
		\textbf{Hidden layer type}  & \textbf{Hidden layers and units as a tuple}  \\
		\hline
		Fully connected input layers  & (512,512)
		\\
		\hline
	LSTM layers	&   (100,100)
		\\
		\hline
	Fully connected output layers	&   (256,256)
		\\
		\hline
	
		
	\end{tabular}
\end{table}

\begin{table}
	\centering
	\caption{The NN structure no LSTM layers, for example in tuple $(x,y)$, length of the tuple indicates the number of hidden layers, and each entity, e.g, $x$, denotes the number of hidden units or dropout ratio.}
	\label{NLSTM_structure}
	\begin{tabular}{ c|c}
		\hline
		\textbf{Hidden layer type}  & \textbf{Hidden layers and units as a tuple}  \\
		\hline
		Fully connected layers  &  (512, 512, 512, 512, 512, 512,\\& 512, 512, 512,
		512,
		256, 128)
		\\
		\hline
		Dropout layers	&   (
		0.4, 0.4, 0.4, 0.4, 0.4, 0.4,\\& 0.4, 0.4, 0.4,
		0.4,
		0.2, 0.2)
		\\
		\hline
		
		
	\end{tabular}
\end{table}
\vspace{-1em}
\subsection{Results Discussions}
In this subsection, we discuss about the simulation results achieved for the following main scenarios:
\begin{enumerate}
	\item \textit{Proposed  LSTM PPO-agent PFR (\textbf{LSTM-PPO})}:
	We propose the on-policy PPO-agent, where the agent's NN layers and hyperparameters are described in Table \ref{Simulation_Sett}, and Table \ref{PPO_hyperparam}, respectively.
	\item \textit{Proposed LSTM SAC-agent PFR (\textbf{LSTM-SAC})}:
	Also, we propose the off-policy SAC agent with the hybrid NN structure shown in Table \ref{Simulation_Sett} as the second approach. The agent's hyperparameters are described in Table \ref{SAC_hyperparam}.
	\item \textit{No-LSTM PPO as a baseline (\textbf{NLSTM-PPO}):} In this baseline, we do not use hybrid layers in the agent's network structure. The NN structure is described in Table \ref{NLSTM_structure}.
\vspace{-0.2em}
\begin{table}
	\centering
	
	\caption{PPO hyperparameters.} 
	\label{PPO_hyperparam}
	\begin{tabular}{ c|c}
		\hline
		\textbf{Hyperparameter}  & \textbf{Value}  \\
		\hline
		Number of epochs  & 25 
		\\
		\hline
		learning rate	&  4e-4 
		\\
		\hline
		Entropy regularization coefficient 	&  1e-2 
		\\
		
		\hline
		Value estimation coefficient 	&  1
		\\
		\hline
		surrogate clip ratio 	&  0.2
		\\
		
		
	\end{tabular}
\end{table}

\begin{table}
	\centering
	
	\caption{SAC hyperparameters.} 
	\label{SAC_hyperparam}
	\begin{tabular}{ c|c}
		\hline
		\textbf{Hyperparameter}  & \textbf{Value}  \\
		\hline
		Number of epochs  & 25 
		\\
		\hline
		learning rate	&  3e-4 
		\\
		\hline
		Reward scale factor 	&  1
		\\
		
		\hline
		target update period  	&  1
		\\
		
		\hline
		target update tau 	&  5e-3
	\end{tabular}
\end{table}
\end{enumerate}
\indent
\vspace{-3em}
\\\indent
	We use multiple approaches and hyperparameters tuning to solve the problem in this paper to get the best outcome. 
	Additionally, we examine our approaches with no discount factor, i.e, $\gamma = 1$, and with considering discount factor, i.e, $\gamma = 0.99$. The motivation is to emphasize the impact of the discount factor on the current action and the most rewarded action. Moreover, we used early stopping method when an agent achieves an acceptable level of accuracy, i.e., when the agent reaches a good performance in all normal, warning, and critical states, to prevent model over-fitting. Approaches and results are discussed as follows:
	\\\indent	
	\subsubsection{Analysis on the Orchestrator's Decisions}
\begin{figure}
		\centering
		\includegraphics[width=0.95\linewidth]{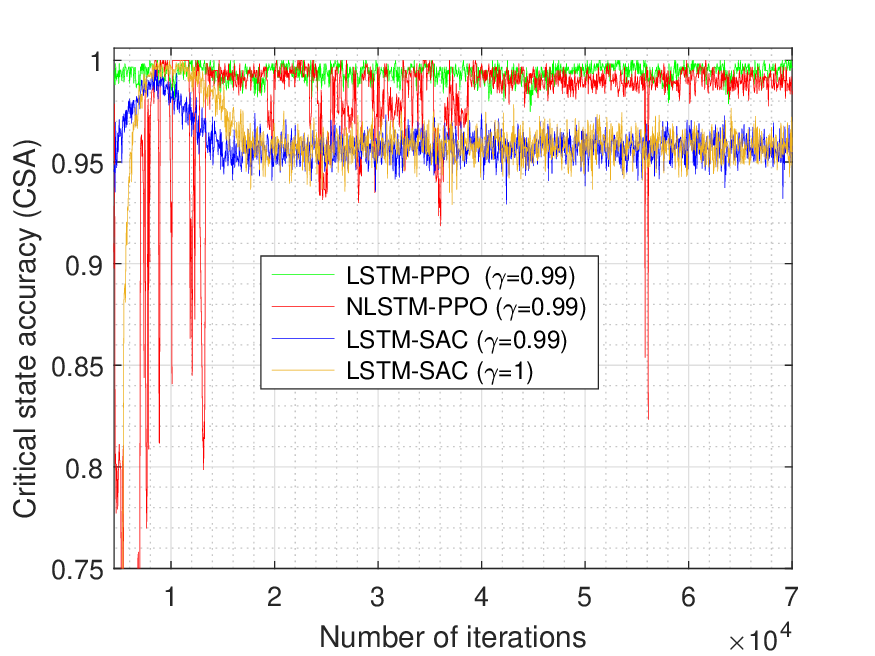}
		\caption{Critical state accuracy comparison for different algorithm under the evolution of time}
		\label{fig:critical-accuracy}
	\end{figure}

\begin{figure}
		\centering
		\includegraphics[width=0.95\linewidth]{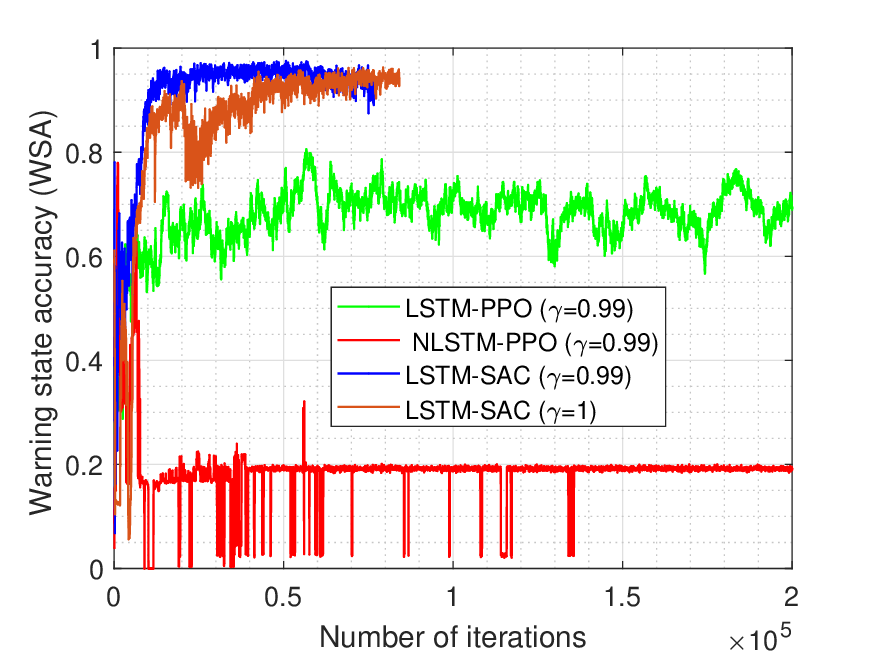}
		\caption{Warning state accuracy comparison for different algorithm under the evolution of time}
		\label{fig:warning-accuracy}
	\end{figure}

\begin{figure}
	\centering
	\includegraphics[width=0.95\linewidth]{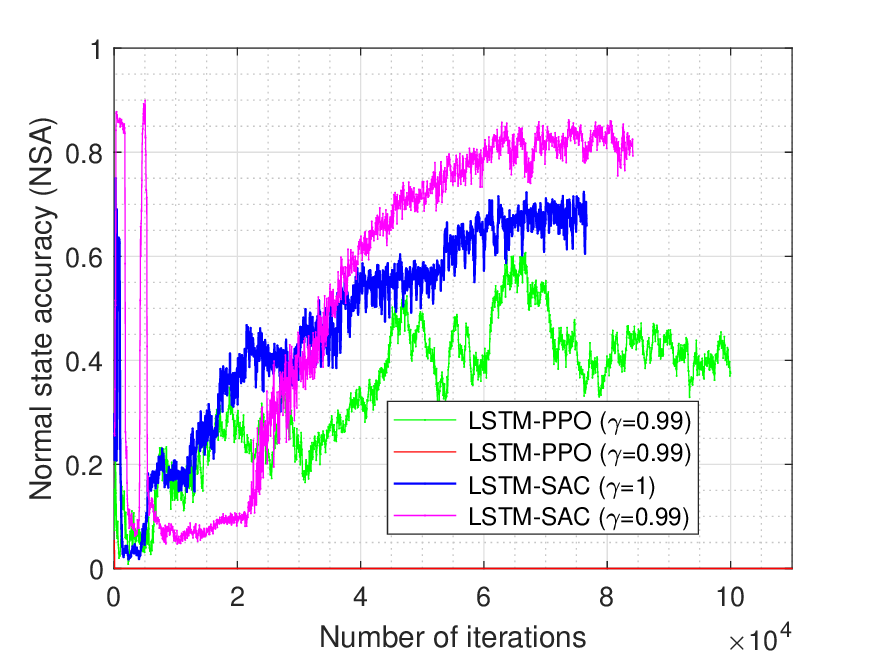}
	\caption{Normal state accuracy comparison for different algorithms under the evolution of time}
	\label{fig:normal-accuracy}
\end{figure}

	In this section, we discuss the agents decision-making in each state of every VNF.
	\\$\bullet~$\textbf{Analysis on Critical State:}
	We define the critical state accuracy as the ratio of detecting critical state and taking desired actions, as follows:
	\begin{align*}
		\text{CSA}\triangleq\frac{\text{Number of taking correct actions in critical state}}{\text{Total number of critical states occurrence}},
	\end{align*}
	    which is shown in Fig. \ref{fig:critical-accuracy}.
	It can be seen from Fig. \ref{fig:critical-accuracy}, LSTM-PPO outperforms LSTM-SAC, meaning that the agent properly understands the critical state and takes the desired actions (as defined before). Even the baseline NLSTM-PPO gets similar results as the LSTM-PPO. PPO is well-known for fast convergence \cite{ppo}, it does also converge faster to high critical state detection accuracy in our model as expected. As mentioned before, the sampled policy of LSTM-PPO and LSTM-SAC in a new environment achieves approximately 99.7\% and 96.6\% CSA, respectively. 
	\\$\bullet~$\textbf{Analysis on Warning State:}
	Fig. \ref{fig:warning-accuracy} depicts the ratio of taking the BP action in the warning state namely warning state accuracy (WSA) over time evolution, defined by
	\begin{align*}
			 \text{WSA}\triangleq\frac{\text{Number of taking correct actions in warning state}}{\text{Total number of warning states occurrence}}.
	\end{align*}
	As shown in Fig. \ref{fig:warning-accuracy}, the LSTM-SAC agent reaches better WSA compared to LSTM-PPO and prepares the service for possible failures. The baseline NLSTM-PPO shows a very poor functionality on understanding to prepare service for failures and service degradation. Accordingly, it seems that  the LSTM-PPO gets stuck in a local optimal solution, which is a well-known issue for on-policy training. In the new environment analysis, the sampled policy shows similar functionality in the training environment. For further analysis of the agent's impending-failure intuition, we study the sampled policy functionality on warning states consistency. Both agents with hybrid NN structure understand the warning consistency impact on critical state occurrence. For example, when a VNF enters the warning state for the first time, LSTM-PPO and LSTM-SAC take the BP action in approximately 45\% and 65\% of the times it occurs. But, if the VNF stays in the warning state for longer than two steps (leading to the critical state), agents understand the impending-failure notion and take the BP action in more than 97\% of the times it occurs. These promising results show that agents with hybrid NN structure figure out the dynamics of our modeled environment on a model-free training basis. 
	\\$\bullet~$\textbf{Analysis on Normal State:}
	Fig. \ref{fig:normal-accuracy} illustrates the ratio of taking the BR action in the normal state namely normal state accuracy (NSA) over time slots, defined by
		\begin{align*}
					 \text{NSA}\triangleq\frac{\text{Number of taking correct actions in normal state}}{\text{Total number of normal states occurrence}}.
		\end{align*}
	As discussed before, the correct action for this state is to remove the placed backup and failure preparations, which means efficient resource usage, and reducing unnecessary  costs. From the figure, LSTM-SAC clearly operates better on normal states and reduces a reasonable amount of unnecessary cost. In our model, the learning rate emphasizes the importance of taking the right action for all states as they take place. The agent with $\gamma = 1$ operates better, because its reward was not discounted during the progress of steps, and its correct action positive reward has  better impact on training policy. As further explanation, our agent gets nearly 100 times better reward on 
	realizing PFR. Therefore, the trained policy's higher priority is to achieve PFR to get the highest reward, and due to lesser rewards of correct action in the normal states, the BR action has lower priority. The discounted reward encourages this prioritized behavior. The motivation of using  rewards with no discount was to smooth this prioritized behavior and to get better results even in the normal states. The normal states do not directly depend on impending-failure occurrence, i.e., direct normal to critical (failure) transition probability is considered zero \cite{ETSI_fault}. Therefore, the achieved reward by the BR action is independent of achieving PFR rewards. As a result, using no-discount reward in PFR achieves better performance in this state.

\subsubsection{Analysis on PFR}
\begin{figure}
	\centering
	\includegraphics[width=0.95\linewidth]{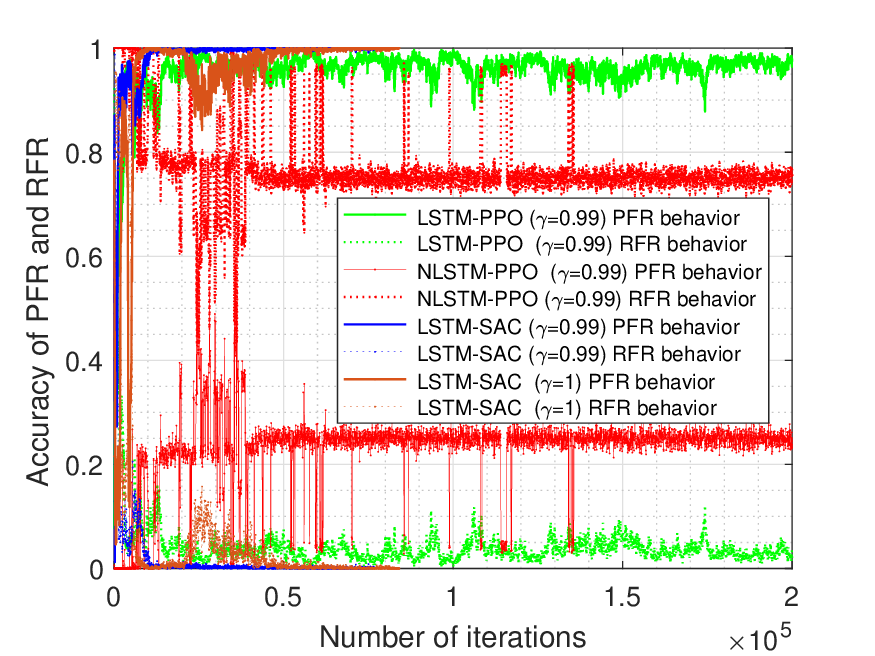}
	\caption{PFR and RFR accuracy  for all algorithms versus the evolution of iterations}
	\label{fig:proreactive-accuracy}
\end{figure}

To give a clear comparison between the PFR and  RFR behavior of our designed agent, first, we define  PFR accuracy and RFR accuracy, respectively, as follows 
\begin{align*}
	\frac{\text{Portion of detected critical states recovered with PFR}}{\text{Number of all detected critical states}},
\end{align*}
and 
\begin{align*}
	\frac{\text{Portion of detected critical states recovered with RFR}}{\text{Number of all detected critical states}}.
\end{align*}
This accuracy  comparison is shown in Fig. \ref{fig:proreactive-accuracy}. 
The figure shows that LSTM-SAC with no discount reward seems to have small fluctuation around $25\times10^3$ iterations, but after $5\times10^4$ iterations, it converges to excellent performance, and approximately  in all times, it manages to do PFR on detected critical states. LSTM-SAC with discounted reward shows a similar performance to the no discount version. Clearly, LSTM-PPO could not manage performance as well as LSTM-SAC, but the results are acceptable. Furthermore, even after $3\times10^5$ iterations, NLSTM-PPO shows a poor performance on figuring out the notion of PFR. Note that, when a VNF is recovered in a proactive manner, it means that the first two steps of the recovery procedure are executed  before failure occurrence, and by modification of synchronization bandwidth, the recovery delay in SLA is not violated. Therefore, the higher PFR percentage implies that most of times failures are detected and fixed, which results in better NFV-based network performance and the user's quality of experience (e.g., by minimizing interruption times for a running services).
\begin{figure}
	\centering
	\includegraphics[width=0.95\linewidth]{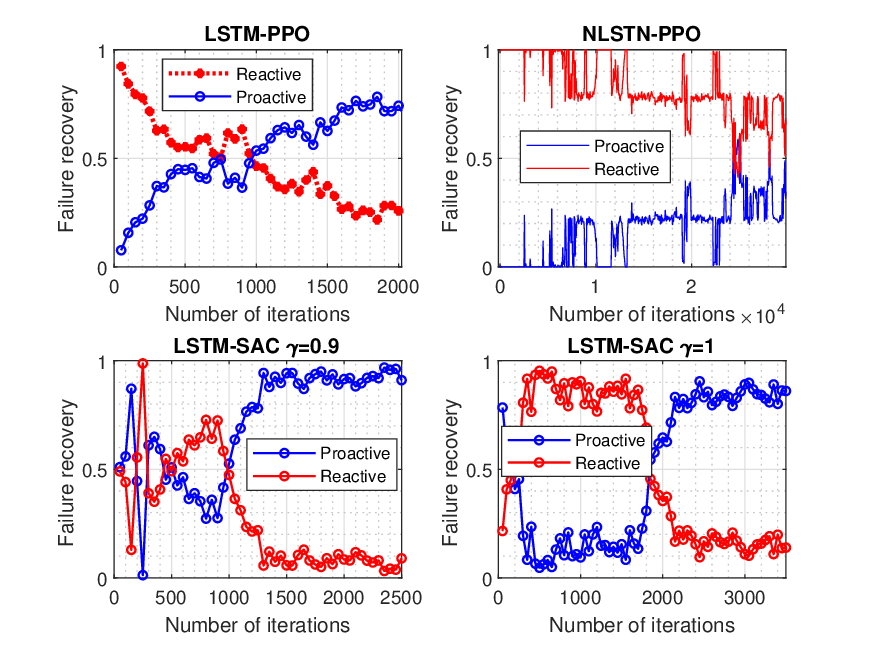}
	\caption{Accuracy of PRF and RFR versus the number of iterations for different algorithms.
	}
	\label{fig:proreactive-accuracy-zoom}
\end{figure}
Moreover, Fig. \ref{fig:proreactive-accuracy-zoom} illustrates how  different agents learn the notion of PFR during time evolution, i.e., the number of iterations. 
As seen from the figure, as  time grows, the agents with hybrid NN tend to perform PFR more than RFR which is the result of  setting appropriate reward function. 
	\subsubsection{Analysis on the Effect of Network Dimension}
\begin{figure}
	\centering
	\includegraphics[width=0.95\linewidth]{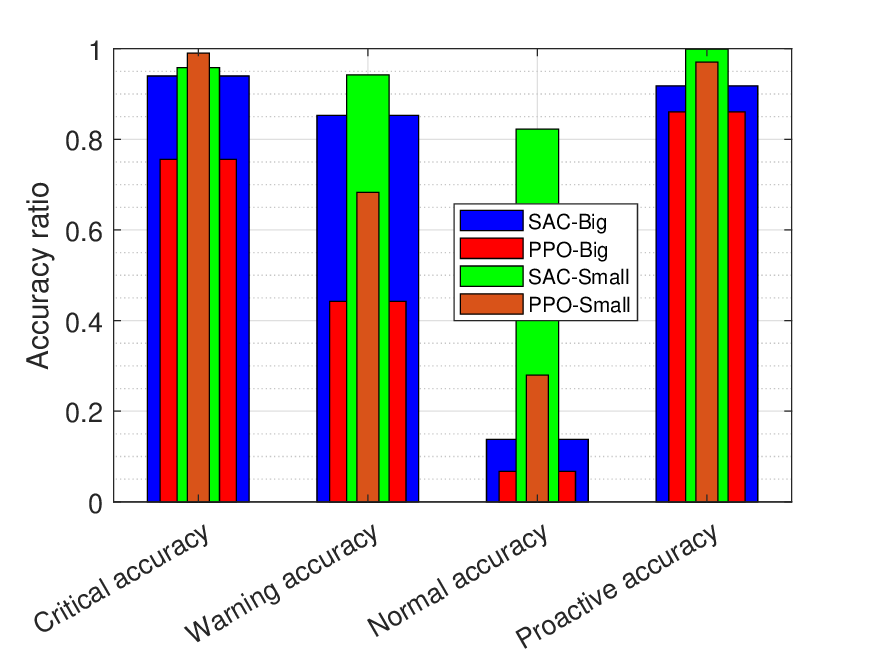}
	\caption{Effect of the network size on the performance of different algorithms. The term "Big" in the figure denotes a larger scale compared to the basic small model, and not a large NFV-based network.}
	\label{fig:bigsmall}
\end{figure}

	In Fig. \ref{fig:bigsmall}, we evaluate the impact of the network dimension, i.e., $K$ and $V$, on the aforementioned accuracy metrics. 
	As shown in Fig. \ref{fig:bigsmall}, our proposed agents has done a better job on smaller network dimension. The reason is that we tried to correct every event in the network simultaneously, and as the network dimension grows, the action space and state space grow too. Therefore, the overall performance degrades. But, the results with LSTM-SAC, show reasonable performance on warning and critical detection, and also reasonable results on PFR are observed. It is worthwhile to mention that as the network dimension grows, more failures could occur along an episode, and therefore, the agent would get more collective reward by completing PFR on failed VNFs.
		As a result, in normal state, a poor functionality is observed due to higher priority of PFR incurred from different rewards corresponding PFR and BR actions.
	 Accordingly, efficient network resource utilization is not guaranteed, i.e., more resources are utilized.
\subsubsection{Summary of Discussion and Insights in DRL Algorithm Design}
	To summarize the workflow, as the first attempt to simulate our proposed ZT-PFR, we used the aforementioned NLSTM-PPO structure. But as discussed, the outcome was not reasonable and the agent could not figure out how to recover a failed VNF in the intended proactive manner. Due to the time-dependent nature of our proposed model and the impending-failure notion, we applied a hybrid NN consisting of LSTM layers, called LSTM-PPO, to capture time-dependent features to cover the impending-failure notion.
	\\\indent
	 The results of LSTM-PPO indicate better performance in figuring out PFR and achieve a reasonable accuracy. But due to insufficient exploration in on-policy PPO methods, the agent gets stuck in a local optimum \cite{sutton2018reinforcement}. However, LSTM-PPO achieves a reasonable level of accuracy (shown in Figs. 4-6), but it does not get better through longer iterations. To achieve better accuracy and solve the exploration problem of on-policy methods, we devise the off-policy method named SAC. The results show a remarkable performance, where LSTM-SAC achieves a better performance in almost all of the metrics. It is worthwhile to mention that the PPO agent performs slightly better on detecting near critical states. In addition, the PPO agent takes less time to converge (approximately half of SAC convergence time). For further analysis, we tried to train the proposed LSTM-SAC and LSTM-PPO on a network with bigger dimensions. LSTM-SAC achieves a better performance. However, because of the  higher dimension and problem complexity \cite{SAC}, the results are not as good as for smaller dimensions.
\vspace{-1em}
\section{Conclusions}\label{conclusion}
 	We proposed a resource-efficient \textit{zero-touch PFR} for stateful VNFs in the context of embedded SFC in an underlying NFV-enabled network.
 	We formulated an optimization problem aiming to minimize a weighted cost including 
 	network resource usage cost and  wrong decision penalty. 
 	As a solution, we customized state-of-the-art DRL-based algorithms such as SAC and PPO. 
 	We adopted a hybrid NN structure consisting of LSTM layers to capture the impending-failure time dependency, resulting in  ZT-PFR performance improvement. We proposed a novel simulated environment considering impending-failure concept, inspired by ETSI \cite{ETSI_fault} and ITU \cite{ITU_fault},  to train and test our DRL agents.
 	  Moreover, we applied the concept  of AoI  to strike a balance between the event and scheduling-based monitoring to guarantee the network's tolerable freshness level. 
Several simulation scenarios are conducted to showcase the efficiency   of our DRL algorithms and  provide a fair comparison with baseline methods.
The results  illustrated that no-discount  LSTM-SAC and LSTM-PPO outperform other algorithms with  remarkable performance in ZT-PFR. However, we remark that  NFV environments has an ever-changing nature. Hence, learning methods for such environments should be in an online fashion with fast training and higher sample efficiency, thus studying such methods could be an interesting research direction.
For future works, we intend to examine our ZT-PFR model in practical  network  environments, extend our model to tackle dynamic and ever-changing NFV environments, and improve  its performance in a larger network dimensions.
\vspace{-1em}



\bibliographystyle{ieeetr}
\bibliography{Library}
\end{document}